\documentclass{aa}  
\usepackage{graphicx}
\usepackage{txfonts}
\usepackage{natbib}

\newcommand{\GILDAS}{\texttt{GILDAS}}

\newcommand{\IRAM}{\textrm{IRAM}}
\newcommand{\IRAMthm}{\textrm{IRAM-30m}}
\newcommand{\PdBI}{\textrm{PdBI}}
\newcommand{\CSO}{\textrm{\CSO}}

\newcommand{\ie} {{\em i.e.}}
\newcommand{\eg} {{\em e.g.}}

\newcommand{\HH}   {\mbox{H$_2$}}           
\newcommand{\DCOp}  {\mbox{DCO$^{+}$}}       
\newcommand{\HCOp}  {\mbox{HCO$^{+}$}}       
\newcommand{\HthCOp}{\mbox{H$^{13}$CO$^{+}$}}
\newcommand{\HCO}  {\mbox{HCO}}         

\newcommand{\CI}   {\mbox{C\small{I}}}   
\newcommand{\Jone}{\mbox{$J$=1--0}}

\newcommand{\Jthr}{\mbox{$J$=3--2}}
\newcommand{\Jfor}{\mbox{$J$=4--3}}

\newcommand{\emm}[1]{\ensuremath{#1}}   
\newcommand{\emr}[1]{\emm{\mathrm{#1}}} 
\newcommand{\unit}[1]{\emm{\, \emr{#1}}}

\newcommand{\mm}  {\unit{mm}}

\newcommand{\pscm}{\unit{cm^{-2}}}

\newcommand{\Kkms}{\unit{K\,km\,s^{-1}}}

\newcommand{\kms}   {\unit{km\,s^{-1}}}

\renewcommand{\deg}{\emm{^\circ}}

\newcommand{\Tas}{\emm{T_\emr{A}^*}}
\newcommand{\Tmb}{\emm{T_\emr{mb}}}
\newcommand{\Beff}{\emm{B_\emr{eff}}}
\newcommand{\Feff}{\emm{F_\emr{eff}}}

\newcommand{\Tk}{\emm{T_\emr{k}}}
\newcommand{\Tex}{\emm{T_\emr{ex}}}

\newcommand{\about}{\emm{\sim}}

\newcommand{\TabObsMaps}{%
  \begin{table*}
    \caption{Observation parameters for the maps shown in Fig.~\ref{fig:maps-pdbi}
      and~\ref{fig:maps-30m} (Fig.~\ref{fig:maps-30m} is available on-line
      only). The projection center of all the maps is
      $\alpha_{2000} = 05^h40^m54.27^s$, $\delta_{2000} = -02\deg 28'
      00''$.}
    \begin{center}
      {\tiny
        \begin{tabular}{ccrclccccccr}
          \hline \hline
          Molecule & Transition & Frequency  & Instrument & Config. & Beam   & PA     & Vel. Resol. & Int. Time$^{a}$ & T$_{sys}$ & Noise$^{b}$ & \multicolumn{1}{c}{Obs. date} \\
                   &            & GHz        &            &         & arcsec & $\deg$ & \kms{}      & hours         & K       & K           & \\
          \hline
          \HthCOp{} & $ 1 - 0$                          & 86.754288 & PdBI & C \& D & $ 6.76 \times 4.65$ & 13 & 0.2 & 6.5 & 150 & 0.10 & 2006-2007 \\
          HCO     & $1_{0,1}\,3/2, 2 - 0_{0,0}\,1/2,1$ & 86.670760 & PdBI & C \& D & $ 6.69 \times 4.39$ & 16 & 0.2 & 6.5 & 150 & 0.09 & 2006-2007 \\
          CCH     &  $1,3/2\,(2) - 0,1/2\,(1)$         & 87.316925 & PdBI & C \& D & $ 7.24 \times 4.99$ & 54 & 0.2 & 6.9 & 130 & 0.07 & 2002-2003 \\
         \hline
        \end{tabular}}
    \end{center}
    $^{a}$ On-source time computed as if the source were always observed with 6 antennae.
    $^{b}$ The noise values quoted here are the noises at the mosaic phase center
    (Mosaic noise is inhomogeneous due to primary beam correction; it 
    steeply increases at the mosaic edges).

 \begin{center}
      {\tiny
        \begin{tabular}{ccrlcccrclccr}
          \hline \hline
          Molecule & Transition & Frequency  & Instrument & \# Pix. & \Feff{} & \Beff{} & Resol. & Resol. & Int. Time$^{a}$ & T$_{sys}$ & Noise   & Obs. date \\
                   &            & GHz        &            &         &         &         & arcsec & \kms{} & hours           &    K    &  mK       &   \\
          \hline
          \HthCOp{} & \Jone{}                      & 86.754288 & 30m/AB100 & 2 & 0.95 & 0.78 & 28.4 & 0.2 & 2.6/5.0  & 133 & 69 & 2006-2007 \\
          HCO & $1_{0,1}\,3/2,2 - 0_{0,0}\,1/2,1 $ & 86.670760 & 30m/AB100 & 2 & 0.95 & 0.78 & 29.9 & 0.2 & 2.6/5.0  & 133 & 63 & 2006-2007 \\
          HCO & $1_{0,1}\,3/2,1 - 0_{0,0}\,1/2,0 $ & 86.708360 & 30m/AB100 & 2 & 0.95 & 0.78 & 29.9 & 0.2 & 2.6/5.0  & 133 & 63 & 2006-2007 \\
          HCO & $1_{0,1}\,1/2,1 - 0_{0,0}\,1/2,1 $ & 86.777460 & 30m/AB100 & 2 & 0.95 & 0.78 & 29.9 & 0.2 & 2.6/5.0  & 133 & 66 & 2006-2007 \\
          HCO & $1_{0,1}\,1/2,0 - 0_{0,0}\,1/2,1 $ & 86.805780 & 30m/AB100 & 2 & 0.95 & 0.78 & 29.9 & 0.2 & 2.6/5.0  & 133 & 66 & 2006-2007 \\
          \hline
        \end{tabular}}
    \end{center}
 $^{a}$ Two values are given for the integration time: the on-source
    time and the telescope time. \\
    \label{tab:obs:maps}
  \end{table*}}

\newcommand{\TabObsTracks}{%
  \begin{table*}
    \caption{Observation parameters for the HCO deep integrations shown in
      Fig.~\ref{fig:maps-pdbi}. Associated transitions can be found in
      Table~\ref{tab:spec}.  The RA and Dec offsets are computed with
      reference to $\alpha_{2000} = 05^h40^m54.27^s$, $\delta_{2000} =
      -02\deg 28' 00''$. The positions are also given in the coordinate
      system used to display the maps in Fig.~\ref{fig:maps-pdbi} 
      and~\ref{fig:maps-30m}. In this coordinate system, maps are 
      rotated by 14\deg{} counter--clockwise around the projection center,
      located at $(\delta x,\delta y)$ = $(20'',0'')$, to bring the 
      illuminated star direction in the horizontal direction and the 
      horizontal zero has been set at the PDR edge.}
    \begin{center}
      {\tiny
        \begin{tabular}{ccc}
          \hline
          \hline
          Position name & $(\delta$RA,$\delta$Dec) & $(\delta x,\delta y)$ \\ 
                        &         arcsec           &        arcsec         \\ 
          \hline
          ``\DCOp{} peak'' & $(20'',22'')$ & $(44.7'',16.5'')$\\
              ``HCO peak'' & $(-5,0'')$    & $(15.1'', 1.2'')$\\
          \hline 
        \end{tabular}}
    \end{center}
    \begin{center}
      {\tiny
        \begin{tabular}{crlcccrclccc}
          \hline \hline
          Position & Frequency  & Line area$^{a}$ & Instrument & \Feff{} & \Beff{} & Resol. & Resol. & Int. Time$^{b}$ & T$_{sys}$ & Noise   & Obs. date \\
                   & GHz        &   \Kkms{}       &            &         &         & arcsec & \kms{} & hours           &    K      &  mK     &   \\
          \hline
          ``\DCOp{} peak'' &  86.670760  & $0.23\pm0.009$ & 30m/B100 & 0.95 & 0.78 & 28.4 & 0.27 & 0.75/1.5 & 134 & 11 & 2008 \\
                           &  86.708360  & $0.12\pm0.009$ & 30m/B100 & 0.95 & 0.78 & 28.4 & 0.27 & 0.75/1.5 & 134 & 11 & 2008 \\
              ``HCO peak'' &  86.670760  & $0.52\pm0.008$ & 30m/B100 & 0.95 & 0.78 & 28.4 & 0.27 & 0.75/1.5 & 127 & 10 & 2008 \\
                           &  86.708360  & $0.31\pm0.007$ & 30m/B100 & 0.95 & 0.78 & 28.4 & 0.27 & 0.75/1.5 & 127 & 10 & 2008 \\
                           & 173.3773770 & $0.47\pm0.023$ & 30m/C150 & 0.93 & 0.65 & 14.2 & 0.067 & 2.0/4.0 & 667 & 66 & 2008 \\
                           & 173.4060816 & $0.26\pm0.018$ & 30m/C150 & 0.93 & 0.65 & 14.2 & 0.067 & 2.0/4.0 & 667 & 66 & 2008 \\
                           & 173.4430648 & $0.23\pm0.020$ & 30m/C150 & 0.93 & 0.65 & 14.2 & 0.067 & 2.0/4.0 & 667 & 66 & 2008 \\
                           & 260.0603290 & $0.16\pm0.019$ & 30m/C270 & 0.88 & 0.46 & 9.5 & 0.18 & 3.0/6.0 & 740 & 59 & 2008 \\
                           & 260.0821920 & $0.14\pm0.020$ & 30m/C270 & 0.88 & 0.46 & 9.5 & 0.18 & 3.0/6.0 & 740 & 59 & 2008 \\
                           & 260.1335860 & $0.12\pm0.017$ & 30m/C270 & 0.88 & 0.46 & 9.5 & 0.18 & 3.0/6.0 & 740 & 59 & 2008 \\
                           & 260.1557690 & $0.06\pm0.016$ & 30m/C270 & 0.88 & 0.46 & 9.5 & 0.18 & 3.0/6.0 & 740 & 59 & 2008 \\
          \hline
        \end{tabular}}
    \end{center}
    $^{a}$ Values obtained from Gaussian fits performed on the spectra using 
    the main beam temperature scale.
    $^{b}$ Two values are given for the integration time: the on-source
    time and the telescope time. \\
    \label{tab:obs:tracks}
  \end{table*}}

\newcommand{\TabTrans}{%
  \begin{table}
    \caption{Einstein coefficients and upper level energies.}
    \begin{center}
      \begin{tabular}{lcrcr} 
        \hline \hline
        Molecule  & Transition            & Frequency &       $A_{ij}$          &  $E_\emr{up}$       \\
                  &  $J,F-J',F'$          &       GHz & (s$^{-1}$)         &      (K)             \\
        \hline
        HCO    &  $1_{01}-0_{00}$\\
               & $3/2,2-1/2,1$ &   86.670760 &  $4.69 \times 10^{-6}$   &  4.2 \\
               & $3/2,1-1/2,0$ &   86.708360 &  $4.60 \times 10^{-6}$   &  4.2 \\
               & $1/2,1,1/2,0$ &   86.777460 &  $4.61 \times 10^{-6}$   &  4.2 \\  
               & $1/2,0-1/2,1$ &   86.805780 &  $4.71 \times 10^{-6}$   &  4.2 \\
               &$2_{02}-1_{01}$\\
               & $5/2,3-3/2,2$ & 173.3773770 & $4.51 \times 10^{-5}$   &  12.5 \\
               & $5/2,2-3/2,1$ & 173.4060816 & $4.43 \times 10^{-5}$   &  12.5 \\ 
               & $3/2,2-1/2,1$ & 173.4430648 & $3.39 \times 10^{-5}$   &  12.5 \\  
               & $3_{03}-2_{02}$\\
               & $7/2,4-5/2,3$ & 260.0603290 & $1.63 \times 10^{-4}$   &  25.0  \\
               & $7/2,3-5/2,2$ & 260.0821920 & $1.61 \times 10^{-4}$   &  25.0  \\
               & $5/2,3-3/2,2$ & 260.1335860 & $1.45 \times 10^{-4}$   &  25.0  \\
               & $5/2,2-3/2,1$ & 260.1557690 & $1.37 \times 10^{-4}$   &  25.0  \\
        \hline
        \HthCOp{} & \Jone{}    &   86.754288 & $3.2 \times 10^{-5}$    &   4.2   \\
                  & \Jthr{}    &   260.2553390      & $1.3 \times 10^{-3}$    &  25.0   \\
        \hline     
 \end{tabular}
 \label{tab:spec}
\end{center} 
The line frequencies and intensities were extracted from the JPL \cite{pic98}
and CDMS \cite{mul01,mul05} molecular spectroscopy data bases for HCO and 
\HthCOp{} respectively.
\end{table}}

\newcommand{\TabCol}{%
  \begin{table}
    \caption{Inferred column densities and abundances with respect to
      molecular hydrogen, \eg{} $\chi(\mbox{X})=N(\mbox{X})/N(\HH)$.}
      \begin{tabular}{lccc} 
        \hline \hline
        Molecule      & Method & HCO peak    &  DCO$^+$ peak \\
\hline
$N$(H$_2$)\,[\pscm]   & 1.2\,mm cont.        & $1.9 \times 10^{22}$  &  $2.9 \times 10^{22}$\\\hline
$N$(\HCO)\,[\pscm]    &  \Tex{} = 5 K      & $3.2 \times 10^{13}$  &  $4.6 \times 10^{12}$\\
$N$(\HthCOp)\,[\pscm] &  Full excitation     & $5.8 \times 10^{11}$  &  $5.0 \times 10^{12}$ $^*$\\
$N$(\HCOp)\,[\pscm]   & $^{12}$C/$^{13}$C=60 & $3.5 \times 10^{13}$  &  $3.0 \times 10^{14}$\\\hline
$\chi$(\HCO)          &                      & $1.7 \times 10^{-9}$  &  $1.6 \times 10^{-10}$\,$^{\dagger}$\\
$\chi$(\HthCOp)       &                      & $3.1 \times 10^{-11}$ &  $1.7 \times 10^{-10}$\\
$\chi$(\HCOp)         &                      & $1.8 \times 10^{-9}$  &  $1.0 \times 10^{-8}$\\
\hline
\end{tabular}
$^*$ Pety et al. (2007a)
$^{\dagger}$\,$1.7 \times 10^{-9}$ if HCO arises only from the cloud surface (A$_V\simeq3$).
\label{tab:abund}
\end{table}}

\begin{document}
\title{HCO mapping of the  Horsehead : \\
 Tracing the illuminated dense molecular cloud surfaces\thanks{Based on
    observations obtained with the IRAM Plateau de Bure interferometer and
    30~m telescope. IRAM is supported by INSU/CNRS (France), MPG (Germany),
    and IGN (Spain).}}

\author{Maryvonne Gerin\inst{1}%
\and Javier R. Goicoechea \inst{1}\thanks{Current address: 
Departamento de Astrof\'{\i}sica. Universidad Complutense de Madrid, Spain.}%
\and Jerome Pety \inst{2,1}%
\and Pierre Hily-Blant \inst{3}}

\offprints{\email{maryvonne.gerin@lra.ens.fr}}

\institute{
LERMA--LRA, UMR 8112, CNRS, Observatoire de Paris and Ecole Normale
Sup\'erieure, 24 Rue Lhomond, 75231 Paris, France.\\
\email{maryvonne.gerin@lra.ens.fr, jrgoicoechea@fis.ucm.es}%
\and IRAM, 300 rue de la Piscine, 38406 Grenoble cedex, France.\\
\email{pety@iram.fr}
\and
Laboratoire d'Astrophysique, Observatoire de Grenoble, BP 53, 38041 Grenoble
 Cedex 09, France. \\ \email{pierre.hilyblant@obs.ujf-grenoble.fr}}

\date{Received September 2008 ; accepted xxx 2008}

 \abstract
{Far-UV photons strongly affect the physical and chemical state of molecular
gas in the vicinity of young massive stars.}
{Finding molecular tracers of the presence of FUV radiation fields in 
the millimeter wavelength domain is desirable, because IR diagnostics 
(PAHs for instance) are not easily accessible towards high extinction 
line-of-sights. Furthermore, 
gas phase diagnostics provide information on the velocity fields.}
{We have obtained maps of the HCO and H$^{13}$CO$^+$ ground state lines 
towards  the Horsehead edge at $5''$ angular resolution with a
    combination of  Plateau de Bure Interferometer (PdBI) and
    the IRAM-30m telescope observations. These maps have been 
complemented with IRAM-30m observations of several excited transitions 
at two    different positions.}
{Bright formyl radical emission delineates the illuminated edge of
  the nebula, with a faint emission remaining towards the
  shielded molecular core. Viewed from the illuminated star, the HCO
  emission almost coincides with the PAH and CCH emission. 
  HCO reaches a similar abundance than HCO$^+$ in the PDR
  ($\simeq$1-2\,$\times10^{-9}$ with respect to H$_2$).  To our
    knowledge, this is the highest HCO abundance ever measured.
  Pure gas-phase chemistry models fail to reproduce the observed
    HCO abundance by $\sim$2 orders of magnitude, except if reactions of
    atomic oxygen with carbon radicals abundant in the PDR (i.e., CH$_2$)
    play a significant role in the HCO formation. 
Alternatively, HCO could be produced
in the PDR by non-thermal processes such as photo-processing of 
ice mantles and subsequent photo-desorption of either HCO or H$_2$CO,  and
  further gas phase photodissociation. }
{The measured HCO/H$^{13}$CO$^+$ abundance ratio is large towards the PDR
  ($\simeq$50), and much lower toward the gas shielded from FUV radiation
  ($\lesssim$1). We propose that high HCO abundances ($\gtrsim$10$^{-10}$)
  together with large HCO/H$^{13}$CO$^+$ abundance ratios ($\gtrsim$1) are
  sensitive diagnostics of the presence of active photochemistry induced
  by  FUV radiation.}


\maketitle

\TabObsMaps{} %

\TabObsTracks{} %

\section{Introduction}
\label{sec:introduction}

Photodissociation region (PDR) models are used to understand the evolution
of far-UV (FUV; $h\nu$\,$<$13.6\,eV) illuminated matter both in our Galaxy
and in external galaxies.  These sophisticated models have been benchmarked
recently \cite{roll07} and are continuously upgraded
\citep[e.g.,][]{goi07,gon08}.  Given the large number of physical and
chemical processes included in such models, it is necessary to build
reference data sets, which can be used to test the predictive
accuracy of models.
Our team has contributed to this goal by providing a series of
high resolution interferometric observations of the Horsehead nebula
\citep[see][for a summary]{pet07b}.
Indeed, this source is particularly well suited
because of its favorable orientation and geometry, and its moderate
distance~\citep[$\sim$400\,pc;][]{hab05}. We have previously studied the
carbon~\citep[]{tey04,pet05} and sulfur chemistry~\citep{goi06}, and
detected the presence of a cold dense core, with active deuterium
fractionation~\citep{pet07}.

The formyl radical, HCO, is known to be present in the interstellar medium
since the late 70's \cite{sny76}.  Snyder et al. (1985) give a detailed
description of the HCO structure and discuss the energy diagram for the
lowest energy levels.  HCO is a bent triatomic asymmetric top with an
unpaired electron.  $a$-type and $b$-type transitions are allowed,
with a stronger dipole moment (1.36 Debye) for the $a$-type
transitions \cite{lan77}, which are therefore more easily detectable. The
strongest HCO ground state transitions lie at 86.670, 86.708, 86.777 and
86.805~GHz, very close to the ground state transition of H$^{13}$CO$^+$ and
to the first rotationally excited SiO line.  Therefore HCO can be
observed simultaneously with SiO and \HthCOp. HCO ground state lines have 
been  detected in the Orion Bar as well as in the dense PDRs NGC~2023, 
NGC~7023 and S~140 \cite{sch01}. From limited mapping, 
they have shown that HCO is sharply
peaked in the Orion Bar PDR, confirming earlier suggestions that HCO is a
tracer of the cloud illuminated interfaces \cite{jon80}.
Garc\'{\i}a-Burillo et al.
(2002) have mapped HCO and H$^{13}$CO$^+$ in the nearby galaxy M82. HCO, CO
and the ionized gas present a nested ring morphology, with the
 HCO peaks being located further out compared to CO and the ring 
of {\sc H\,ii}  regions.
The chemistry of HCO is not well understood.
Schilke et al. (2001) concluded that it is extremely difficult 
to understand the observed
HCO abundance in PDRs with gas phase chemistry alone. As a possible
  way out, they tested the production of HCO by the photodissociation of
  formaldehyde.  In this model, H$_2$CO is produced in grain mantles, and
released by non-thermal photo-desorption in the gas phase in the PDR. 
However, even with this favorable hypothesis, the model can not 
reproduce the abundance and spatial distribution of HCO because 
the photo-production is most efficient at an optical
depth of a few magnitudes where the photodissociation becomes less effective.

In this paper, we present maps of the formyl radical ground state 
lines at high   angular resolution towards the Horsehead nebula, and 
the detection of higher energy  level transitions towards two 
particular lines of sights, one in the PDR
  region and the other in the associated dense core. These observations 
enable us to accurately study the HCO spatial distribution and 
abundance. We present the observations and data reduction in 
section \ref{sec:obs}, while
the results and HCO abundance are given in section \ref{sec:results}, and
the discussion of HCO chemistry and PDR modeling is given in section
\ref{sec:models}.

\section{Observations and data reduction}
\label{sec:obs}

Tables \ref{tab:obs:maps} and \ref{tab:obs:tracks} summarize the
observation parameters for the data obtained with the \IRAM{} \PdBI{} and
30m telescopes.  The HCO ground state lines were observed simultaneously
with \HthCOp and SiO.  Frequency-switched, on-the-fly maps of the
\HthCOp{}~\Jone{} and HCO ground state lines 
(see  Fig.~\ref{fig:maps-30m}), obtained at the \IRAMthm{}
using the A100 and B100 3mm receivers ($\about 7\mm$ of water vapor) were used
to produce the short-spacings needed to complement a 7-field mosaic acquired
with the 6 \PdBI{} antennae in the CD configuration (baseline lengths from
24 to 176~m). The whole \PdBI{} data set will be comprehensively described
in a forthcoming paper studying the fractional ionization across the
Horsehead edge (Goicoechea et al. 2009, in prep.). The CCH data 
shown in Fig.~\ref{fig:maps-pdbi} have been extensively 
described in Pety et al. 2005.
The high resolution HCO $1_{0,1}-0_{0,0}$ data are complemented by
observations of the $2_{0,2}-1_{0,1}$ and $3_{0,3}-2_{0,2}$ multiplets with
the \IRAM{} 30m telescope centered on the PDR and the dense core. To obtain
those deep integration spectra, we used the position switching observing mode.
The on-off cycle duration was 1 minute and the off-position offsets were
$(\delta RA, \delta Dec) = (-100'',0'')$, \ie{} the {\sc{H\,ii}} region
 ionized by  $\sigma$Ori and free of molecular emission. 
Position accuracy is estimated to be about $3''$ for the
30m data and less than $0.5''$ for the PdBI data.

The data processing was done with the \GILDAS{}\footnote{See
  \texttt{http://www.iram.fr/IRAMFR/GILDAS} for more information about the
  \GILDAS{} softwares.} softwares~\citep{pet05b}.  The \IRAMthm{} data were
first calibrated to the \Tas{} scale using the chopper wheel
method~\citep{pen73}, and finally converted to main beam temperatures
(\Tmb{}) using the forward and main beam efficiencies (\Feff{} \& \Beff{})
displayed in Table~\ref{tab:obs:tracks}.  The resulting amplitude accuracy
is \about{} 10\%.  Frequency-switched spectra were folded using the
standard shift-and-add method, after baseline subtraction. The resulting
spectra were finally gridded through convolution by a Gaussian.
Position-switched spectra were co-added before baseline subtraction.
Interferometric data and short-spacing data were merged before imaging and
deconvolution of the mosaic, using standard techniques of
\GILDAS{}~\citep[see \eg{}][for details]{pet05}.

\section{Results and Discussion}
\label{sec:results}

\subsection{Spatial distribution}

\begin{figure*}[t]
  \centering %
  \includegraphics[height=\hsize{},angle=270]{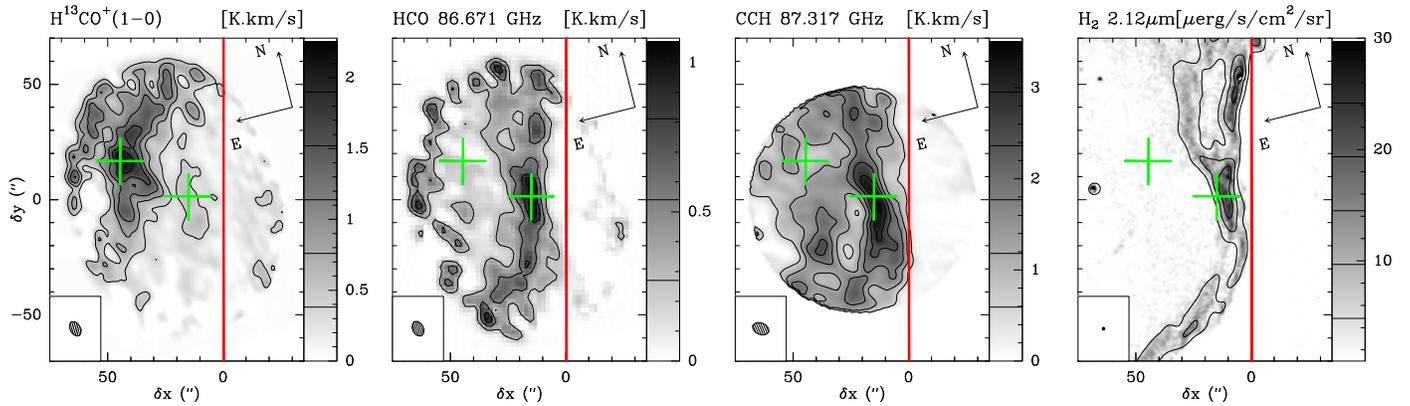} %
  \caption{High angular resolution maps of the integrated intensity of
      \HthCOp{}, HCO,  CCH and vibrationally excited H$_2$ emission. 
\HthCOp{} and HCO have been observed
      simultaneously, both with the IRAM-30m and IRAM-PdBI. Maps have been
      rotated by 14\deg{} counter--clockwise around the projection center,
      located at $(\delta x,\delta y)$ = $(20'',0'')$, to bring the
      illuminated star direction in the horizontal direction and the
      horizontal zero has been set at the PDR edge. The emission of all
      lines is integrated between 10.1 and 11.1 \kms. Displayed integrated
      intensities are expressed in the main beam temperature scale. 
    Contour levels are displayed on the grey scale lookup tables. The
      red vertical line shows the PDR edge and the green crosses show the
      positions (\DCOp{} and HCO peaks) where deep integrations have
      been performed at IRAM-30m (see Fig.~\ref{fig:mods-30m}). 
The H$_2$ map is taken from \citet{hab05}.}
  \label{fig:maps-pdbi}
\end{figure*}

  Fig.~\ref{fig:maps-pdbi} shows a map of the integrated intensity of
  the strongest HCO line at 86.671 GHz, of the \HthCOp{} \Jone{} line and
  of the strongest CCH line at 87.317 GHz. Fig.~\ref{fig:mods-30m} displays
  high signal-to-noise ratio spectra of several hyperfine components of three
  HCO rotational transitions towards the HCO and the \DCOp{} emission
  peaks.
  
 Most of the formyl radical emission is concentrated in a narrow
  structure, delineating the edge of the Horsehead nebula. Low level
  emission is however detected throughout the nebula, including towards the
  dense core identified by its strong \DCOp{} and \HthCOp{} emission
  \cite{pet07}.  The HCO emission is resolved by our PdBI observations.
  From 2-dimensional Gaussian fits of the image, we estimate that the
  emission width is $\sim$13$\pm$4\arcsec\ in the plane of the sky. 
  The \HthCOp{} emission
  shows a different pattern : most of the signal is associated
with the dense core behind the photodissociation front, and  faint
  \HthCOp{} emission detected in the illuminated edge. 
The CCH emission pattern is less extreme than HCO, but 
shows a similar enhancement in the PDR.
  
 In summary, the morphology of the HCO emission is reminiscent of
  the emission of the PDR tracers, either the PAH emission \citep{abe02} or
  the small hydrocarbons emission, which is strongly enhanced towards the
  PDR~\citep{tey04,pet05}. By contrast, the HCO emission becomes strikingly
  faint where the gas is dense and shielded from FUV radiation. 
These regions are
  associated with bright  DCO$^+$ and \HthCOp{}
  emission~\citep{pet07}.   
Our maps therefore confirm that HCO is a PDR species.

\subsection{Column densities and abundances}

\subsubsection{Radiative transfer models of HCO and \HthCOp{} }

\TabTrans{} %

 Einstein coefficients and upper level energies of the studied
HCO and \HthCOp{} lines are given in Table \ref{tab:spec}.
 As no collisional cross--sections with H$_2$ nor He have been 
calculated for HCO so far, we have computed the HCO column densities 
assuming a single excitation temperature \Tex{} for all transitions. 
Nevertheless our calculation takes into account
thermal, turbulent and opacity broadening as well as the cosmic microwave
background and line opacity \cite{goi06}. 
For \HthCOp{},  detailed  non-local and non-LTE excitation and 
radiative transfer calculations have
been performed using the same approach as in our previous PdBI CS and
C$^{18}$O line analysis (see Appendix in Goicoechea et al. 2006).
\HthCOp{}-H$_2$ collisional rate coefficients were adapted from those of
Flower (1999) for \HCOp{}, and specific \HthCOp{}-electron rates where
kindly provided by Faure \& Tennyson (in prep.).

\subsubsection{Structure of the PDR in HCO and \HthCOp{}}

\begin{figure}[h]
  \centering %
  \includegraphics[width=5cm,angle=270]{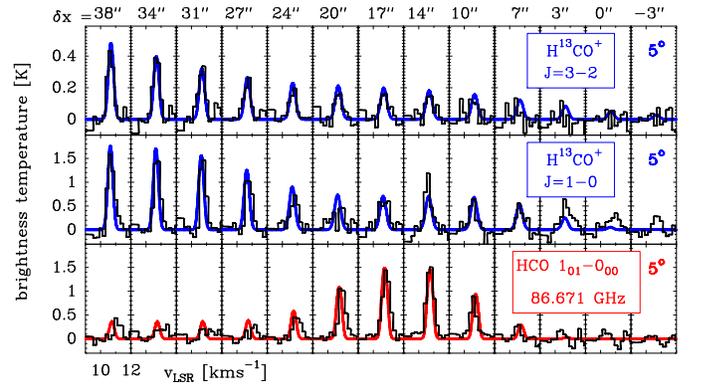} %
   \caption{Observations along a horizontal cut through ``the HCO peak'' (histograms). 
     The \HthCOp{} $J$=1-0 and HCO $1_{01}-0_{00}$ lines were mapped with
     the PdBI at an angular resolution of 6.8$''$, whereas the \HthCOp{}
     $J$=3-2 line was mapped with HERA-30m (and smoothed to a spatial
     resolution of 13.5$''$).  Radiative transfer models of an edge-on
     cloud with a line of sight extinction of A$_V$=20, 
      inclined 5$\deg$ relative to the
     line of sight for HCO (red curve), and \HthCOp{} (blue curves) are
     shown. The single-\Tex{} HCO model assumes a 12$''$ width filament
     with a column density of $3.2 \times 10^{13}$~\pscm{}, while $N$(HCO)
     is $4.6 \times 10^{12}$~\pscm{} behind the filament. The \HthCOp{}
     model assumes a constant density of
     $n$(H$_2$)=5$\times$10$^4$\,cm$^{-3}$ with \Tk{}=60\,K and
     $N$(\HthCOp{})=$5.8 \times 10^{11}$~\pscm{} for $\delta x$\,$<$35$''$;
     and \Tk{}=10\,K and $N$(\HthCOp{})=$7.6 \times 10^{11}$~\pscm{} for
     $\delta x$\,$>$35$''$.  Modeled line profiles have been convolved with
     an appropriate Gaussian beam corresponding to each PdBI synthesized
     beam or 30m main beam resolution.}
  \label{fig:mods-PdBI}
\end{figure}

To get more insight on the spatial variation of the HCO and \HthCOp 
column densities and abundances, we have analyzed a cut 
through the PDR, centered on the ``HCO peak''
 at $\delta y$=0'' (see Fig.~\ref{fig:mods-PdBI}).
The cut clearly shows that HCO is brighter than
  \HthCOp{}  in the PDR and vice-versa in the dense core.
Taking into account the different level degeneracies of both
transitions (a factor 2.4) and the fact that the associated Einstein
coefficients $A_{ij}$ differ by a factor $\sim$8 (due to the different
permanent dipole moments, see Table~\ref{tab:spec}), $N$(\HthCOp{}) must be
significantly lower than $N$(HCO) towards the PDR.

 We modeled the PDR as an edge-on cloud inclined by $\sim$5$\deg$
relative to the line-of-sight. We have chosen a cloud depth of $\sim$0.1\,pc,
which implies an extinction of $A_V$\,$\simeq$20 mag for the considered
densities towards the ``HCO peak''. These parameters are the best
geometrical description of the Horsehead PDR-edge \cite[e.g.,][]{hab05} and
also reproduce the observed 1.2\,mm continuum emission intensity.
The details of this modeling will be presented in Goicoechea et al.~(2009).
In the following, we describe in details
  the determination of the column densities and abundances for two
  particular positions, namely the ``HCO peak'' and the ``DCO$^+$ peak''
  (offsets relative to the map center can be found in
  Table~\ref{tab:obs:tracks}).

\begin{figure*}[ht]
  \centering %
  \includegraphics[width=9cm,angle=270]{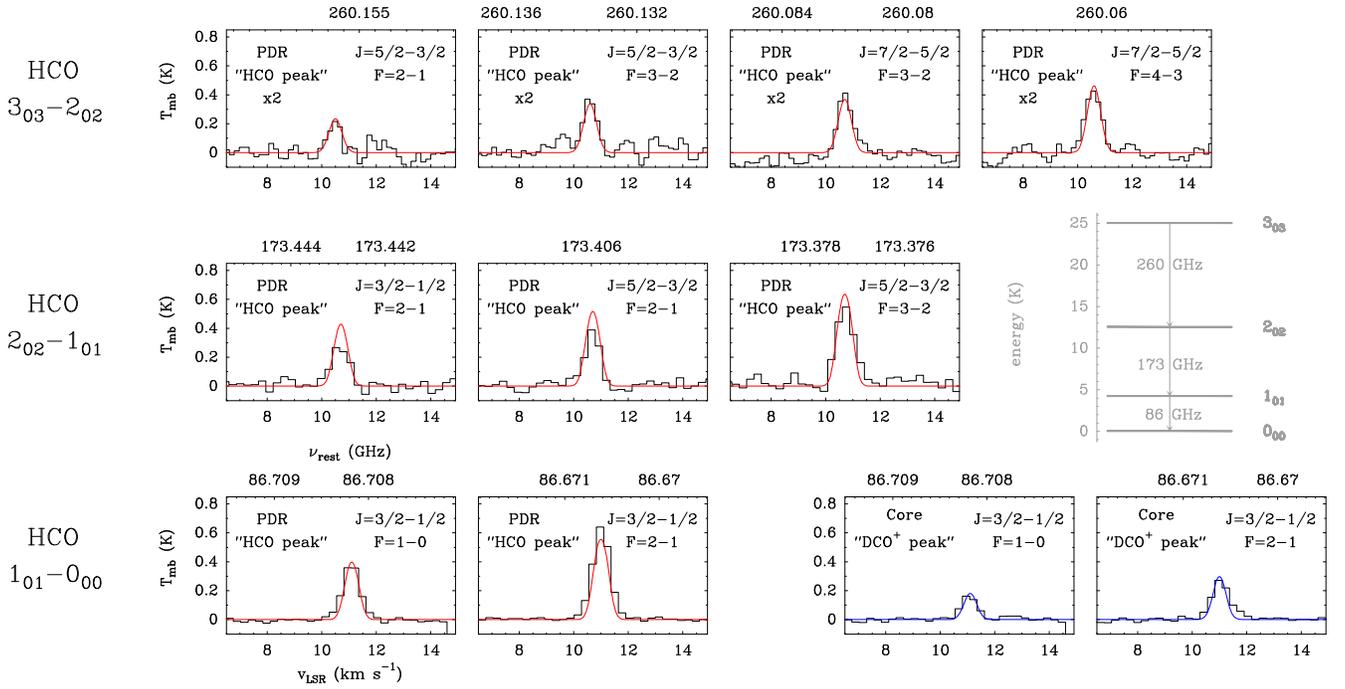} %
   \caption{IRAM-30m observations (histograms) of several HCO hyperfine components
     of the 1$_{01}$-0$_{00}$, 2$_{02}$-1$_{01}$ and 3$_{03}$-2$_{02}$
     rotational transitions towards the PDR (``HCO peak'') and towards the
     dense core (``DCO$^+$ peak'') \cite{pet07}.  Solid lines are
     single-\Tex{} radiative transfer models of the PDR-filament (red
     curves) and line-of-sight cloud surface (blue curves).  A sketch of
     the HCO rotational energy levels is also shown (\textit{right
       corner}). }
  \label{fig:mods-30m}
\end{figure*}

\subsubsection{HCO  column densities}

We used the three detected rotational transitions of HCO (each with
  several hyperfine components, see  Fig.~\ref{fig:mods-30m}) to
  estimate the HCO column densities in the direction of the ``HCO'' peak.
We have taken into account the
varying beam dilution factors of the HCO emission at 
the ``HCO peak'' by modeling the HCO
emission as a Gaussian filament of $\sim$12$''$ width in the
$\delta x$ direction, and infinite in the $\delta y$
direction. The filling factors at 260, 173 and 87 GHz are thus $\sim$0.8,
0.6 and 0.4, respectively.  

A satisfactory fit of the IRAM--30m data
towards the ``HCO peak'' is obtained for \Tex{}\,$\simeq$5~K and a turbulent
velocity dispersion of $\sigma$=0.225$\kms$\,(FWHM$=2.355\times\sigma$).
Line profiles are reproduced  for $N$(HCO) = $3.2 \times
10^{13}$ \pscm{} (see red solid curves in Fig.~\ref{fig:mods-30m}).  The
most intense HCO lines at 86.67 and 173.38~GHz become marginally optically
thick at this column density ($\tau \gtrsim 1$). Therefore, opacity
corrections need to be taken into account.  We checked that the low value
of \Tex{} (subthermal excitation as $\Tk{}$\,$\simeq$60\,K) is consistent
with detailed excitation calculations carried for \HthCOp{} in the PDR which
are described below.

Because the HCO signals are weaker towards the ``\DCOp peak'', we only detected
2 hyperfine components of the $1_{01}-0_{00}$ transition. Assuming extended
emission and the same excitation temperature as for the ``HCO peak'', 5~K, we
fit the observed lines with a column density of $4.6 \times 10^{12}$ \pscm{}
(blue solid lines in Fig.~\ref{fig:mods-30m}). 
Both HCO lines are optically thin at this position.
This simple analysis shows that the HCO column density is $\sim 7$ times
larger at the ``HCO peak'' in the PDR, than towards the dense cold core.

\subsubsection{\HthCOp{} column densities}

Both the \HthCOp{} $J$=3-2 and 1-0 line profiles at the ``HCO
  peak'' are fitted with $n(H_2)$\,$\simeq$5$\times$10$^4$\,cm$^{-3}$, 
\Tk{}\,$\simeq$60\,K
and $e^-$/H\,$\simeq$5$\times$10$^{-5}$ (as predicted by the PDR
models below). The required column density is $N$(\HthCOp{})=$5.8
\times 10^{11}$~\pscm{}. For those conditions, the excitation 
temperature, \Tex{},  of the $J$=3-2 transition varies 
from $\simeq$4 to 6\,K, which supports the
single-\Tex{} models of HCO. Both \HthCOp{} lines are optically thin
towards the ``HCO peak''.
 
The \HthCOp{} line emission towards the ``DCO$^+$ peak'' has been studied
by Pety et al. (2007a).  Both \HthCOp{} lines are moderately optically thick
towards the core, and the \HthCOp{} column density 
is $N$(\HthCOp{})$ \simeq 5.0 \times 10^{12}$~\pscm{}, which 
represents an enhancement of nearly one
order of magnitude relative to the PDR.  According to our 1.2\,mm continuum
map, the extinction towards the core is A$_V$\,$\gtrsim$30 mag
compared to 20 mag in the PDR. The \HthCOp column density
enhancement therefore corresponds to a true abundance enhancement.

\subsubsection{Comparison of HCO and  \HthCOp{} abundances }

\TabCol{}%

Table \ref{tab:abund} summarizes the inferred HCO and \HthCOp{} column
densities and abundances towards the 2 selected positions : 
the ``HCO peak'' in the PDR  and
the ``DCO$^+$ peak'' in the FUV-shielded core. Both species exhibit
strong variations of their column densities and abundances relative to H$_2$
between the PDR and the shielded region.
In the PDR, we found that  both the HCO abundance relative to H$_2$
($\chi$(HCO)$\simeq$1-2\,$\times10^{-9}$) and the HCO/\HthCOp{} column
density ratio ($\approx$50)  are high. These figures are higher than all 
previously published measurements (at
lower angular resolution).  Besides, 
\textbf{the formyl radical and HCO$^+$ reach similar abundances in the
  PDR}.

\begin{figure*}[ht]
  \centering %
  \includegraphics[width=9.9cm,angle=270]{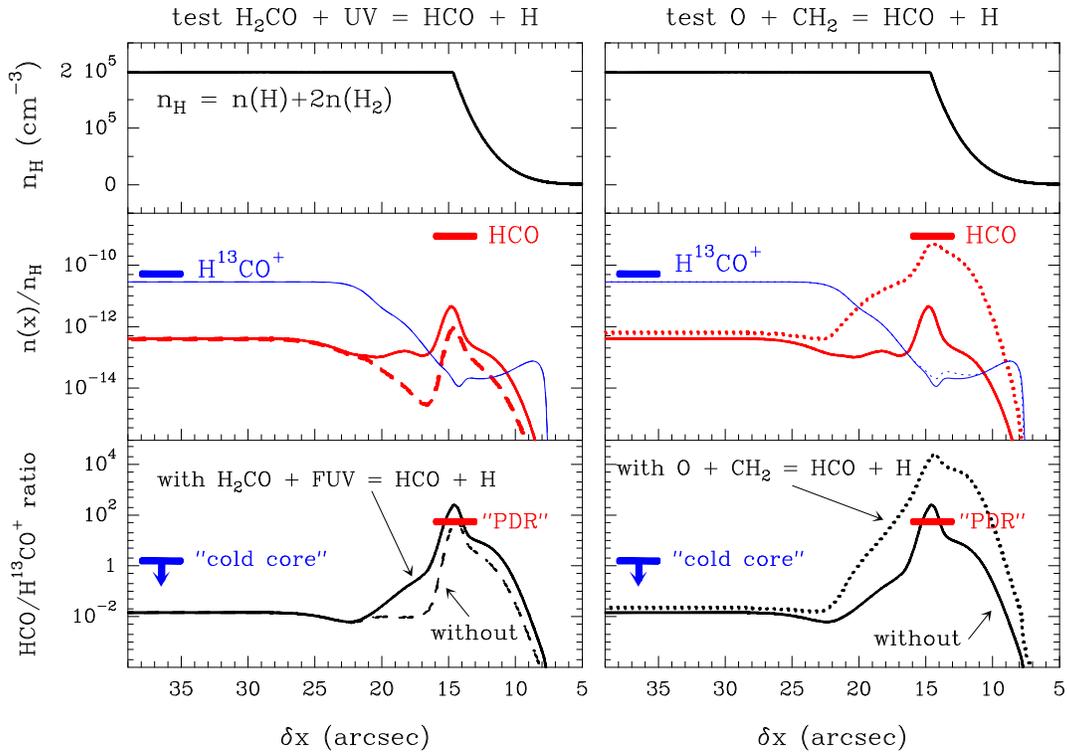} %
   \caption{Photochemical models of a unidimensional PDR.
     \textit{Upper panels} show the density gradient ($n_H=n(H)+2n(H_2)$ in
     cm$^{-3}$) used in the calculation. \textit{Middle panels} show the
     predicted HCO and \HthCOp{} abundances (relative to $n_H$). The
     \HthCOp{} abundance inferred from observations in the cold core (``the
     DCO$^+$ peak'', see the offsets in Table~\ref{tab:obs:tracks}) is 
    shown with
     blue lines.  The HCO abundance inferred from observations in the PDR
     (``the HCO peak'', see the offsets in Table~\ref{tab:obs:tracks}) is shown
     with red lines. \textit{Lower panels} show the HCO/\HthCOp{} abundance
     ratio predicted by the models whereas the HCO/\HthCOp{} column density
     ratio inferred from observations is shown as blue arrows and 
      red lines (for
     the cold core and PDR respectively). Every panel compares two
     different models: \textbf{Left-side models} show a \textit{standard
       chemistry} (dashed curves) versus the same network upgraded with the
     addition of the H$_2$CO + photon $\rightarrow$ HCO + H
     photodissociation (solid curves). \textbf{Right-side models} show the
     previous upgraded \textit{standard} model (solid curves) versus a
     chemistry that adds the O + CH$_2$ $\rightarrow$ HCO + H reaction with
     a rate of 5.01$\times$10$^{-11}$\, cm$^{3}$\,s$^{-1}$ (dotted curves).
     The  inclusion of the O + CH$_2$ reaction has almost no effect on
     H$^{13}$CO$^+$ for the physical conditions prevailing in
     the Horsehead, but triggers an increases of the HCO abundance in the PDR
     by two orders of magnitude. 
}
  \label{fig:mods-chem}
\end{figure*}

The situation is reversed towards the ``DCO$^+$ peak'', \ie{} the 
observed HCO/\HthCOp{} column density ratio is  lower ($\approx$1) 
than towards  the ``HCO peak'' . Nevertheless, while the bulk of 
the observed \HthCOp{} emission arises from
cold and shielded gas, the origin of HCO emission is less clear.
HCO could either $(i)$ coexist with \HthCOp{} or $(ii)$
arise predominantly from the line-of-sight cloud surface.  In the former
case, our observations show that the HCO abundance  drops by one order of
magnitude between the PDR and the dense core environment.  
However, it is possible that the abundance variation is
even more pronounced, if the detected HCO emission arises 
from the line of sight cloud surface. We have estimated the  
depth of the cloud layer, assuming that  HCO keeps the
``PDR abundance'' in this foreground layer : a cloud surface 
layer of A$_V$\,$\simeq$3 (illuminated
by the mean FUV radiation field around the region) also reproduces the
observed HCO lines towards the cold dense core (blue solid lines in
Fig.~\ref{fig:mods-30m}).

In this case, both
the HCO abundance and the HCO/\HthCOp{} abundance ratio in the dense core
itself will be even lower than listed in table \ref{tab:abund}.  
We have tried to discriminate both scenarios by
comparing the HCO $1_{01}-0_{00}$ ($J$=3/2-1/2, $F$=2-1) and H$^{13}$CO$^+$
$J$=1-0 line profiles towards this position.  Both lines have been observed
simultaneously with the IRAM-30m telescope. Because of their very similar
frequencies ($\sim$86.7\,GHz), the beam profile and angular resolution is
effectively the same. In this situation, any difference in the measured
linewidths reflects real differences in the gas kinematics and turbulence
of the regions where the line profiles are formed.  Gaussian
fits of the HCO and \HthCOp{} lines towards ``the DCO$^+$ peak'' provides
line widths of 
$\Delta$v(HCO) = $0.81\pm$0.06\,km\,s$^{-1}$ and  
$\Delta$v(H$^{13}$CO$^+$) = $0.60 \pm$0.01\,km\,s$^{-1}$.  
Therefore, even if the H$^{13}$CO$^+$ $J$=1-0 line are 
slightly broadened by opacity and do not represent the
true line of sight velocity dispersion, HCO lines are  broader at 
the 3$\sigma$ level of confidence.
This remarkable difference supports  scenario $(ii)$ where the H$^{13}$CO$^+$
line emission towards the ``the DCO$^+$ peak'' arises from the quiescent,
cold and dense core, whereas HCO, in the same line of sight, arises
predominantly from the warmer and more turbulent outer cloud layers.
We note that the presence of a foreground layer of more diffuse
material (A$_V$\,$\sim$2  mag) was already introduced by 
Goicoechea et al. ~(2006), to fit the
CS $J$=2-1 scattered line emission. 
The analysis of CO \Jfor \ and \CI \ $^3P_1 - ^3P_0$ maps led
also  \citet{phil06} to propose  the presence of a diffuse envelope, 
with A$_V$\,$\sim$2  mag, and which contributes to roughly the about 
half the mass of the dense filament traced by C$^{18}$O and
the dust continuum emission.  
The hypothesis of a surface layer of 
HCO  is therefore consistent  with previous modeling of molecular
emission of the horsehead. 

We conclude 1) that HCO and HCO$^+$ have similar abundances in the
PDR, and 2) that the HCO abundance drops by at least one order of
magnitude between the dense and warm PDR region and the cold and shielded
DCO$^+$ core.

\section{HCO chemistry}
\label{sec:models}

\subsection{Gas-phase formation: PDR models}

In order to understand the HCO and \HthCOp{} abundances and HCO/\HthCOp{}
column density ratio inferred from observations, we have modeled the steady
state gas phase chemistry in the Horsehead edge.  The density distribution
in the PDR is well represented by a density gradient $n_H(\delta
x)$\,$\propto$\,$\delta x^4$, where $\delta x$ is the distance from the
edge towards the cloud interior and $n_H=n(H)+2n(H_2)$ (see the top
  panels of Fig.~\ref{fig:mods-chem}). The density
reaches a constant $n_H$ value of 2$\times$10$^5$\,cm$^{-3}$ in an
equivalent length of $\sim$10$''$ \cite{hab05,goi06}.  The cloud edge is
illuminated by a FUV field 60 times the mean interstellar radiation field
($G_0 = 60$ in Draine units).  
We used the \texttt{Meudon PDR code}\footnote{Publicly
  available at \texttt{http://aristote.obspm.fr/MIS/}}, a photochemical
model of a unidimensional PDR~\citep[see][for a detailed
description]{jlb93,flp06,goi07}. Our standard chemical network is based on
a modified version of \textit{Ohio State University (osu)} gas--phase
network, updated for photochemical studies \citep[see][]{goi06}.  It also
includes $^{13}$C fractionation reactions \cite{Gra82} and specific
computation of the $^{13}$CO photodissociation rate as a function of depth.
The ionization rate due to cosmic rays in the models is
$\zeta$=5$\times$10$^{-17}$\,s$^{-1}$.  Following our previous works, we
chose the following elemental gas phase abundances: He/H=0.1,
O/H=3$\times$10$^{-4}$, C/H=1.4$\times$10$^{-4}$, N/H=8$\times$10$^{-5}$,
S/H=3.5$\times$10$^{-6}$, $^{13}$C/H=2.3$\times$10$^{-6}$,
Si/H=1.7$\times$10$^{-8}$ and Fe/H=1.0$\times$10$^{-9}$.

In Fig.~\ref{fig:mods-chem}, we investigate the main gas-phase formation
routes for HCO in a series of models "testing" different pathways leading to
 the formation of HCO. HCO and \HthCOp{} predictions are shown in
Figure~\ref{fig:mods-chem} (middle panels). As a first result, note that in
all models the HCO abundance peaks near the cloud surface at
A$_V$\,$\simeq$1.5 ($\delta x$\,$\simeq$14$''$) where the ionization
fraction is high ($e^-$/H$\sim$5$\times$10$^{-5}$). Due to the low
abundance of \textit{metals} in the model (as represented by the low
abundance of Fe), the ionization fraction in the shielded regions is low
(e$^-$/H  $\lesssim$10$^{-8}$), and therefore the \HthCOp{} predictions
matches the observed values (Goicoechea et al. 2009).  Besides, a low
metalicity reduces the efficiency of charge exchange reactions of HCO$^+$
with metals, e.g.,
\begin{equation}
\emr{Fe + HCO^+ \rightarrow HCO + Fe^+}
\label{eq-fe}
\end{equation}
which are the main gas-phase formation route of HCO in the FUV-shielded gas
in our models. Hence, the HCO abundance remains low inside the core.
Nevertheless, despite that such models do reproduce the observed HCO
distribution, which clearly peaks at the PDR position, the predicted
absolute HCO abundances can vary by orders of magnitude depending of the
dominant formation route.

 In our \textit{standard} model (left-side models : dashed curves), 
the formation of HCO in the PDR is dominated
by the dissociative  recombination of H$_2$CO$^+$, while its destruction is 
dominated by photodissociation.
Even if the predicted HCO/\HthCOp{} abundance ratio 
satisfactorily reproduces the value inferred
from observations, the predicted HCO abundance peak is 
$\sim$3 orders of magnitude lower than observed.
In order to increase the gas-phase formation of the HCO in the PDR we have
added  a new channel in the
photodissociation of formaldehyde, the production HCO, 
in addition to the normal channel producing CO :
\begin{equation}
\emr{H_2CO + photon \rightarrow HCO + H}
\label{eq-h2co}
\end{equation}
This channel is generally not included in standard chemical networks but
very likely exists \cite{tro07,yin07}. We included this process
  with an unattenuated photodissociation rate of 
$\kappa_{diss}$(H$_2$CO)=10$^{-9}$\,s$^{-1}$ and a
depth dependence given by exp($-$1.74\,A$_V$).  This is the same rate 
as the one given
by van Dishoeck (1988) for the photodissociation of H$_2$CO producing CO,
which  is explicitly calculated for the Draine (1978) radiation
field.  Model results are shown in Fig.~\ref{fig:mods-chem}
(left-side models: solid curves).  The inclusion of
Reaction~\ref{eq-h2co}, which becomes the dominant HCO formation
route, increases the HCO abundance in the PDR by a factor $\sim$5. 
But the HCO production rate is still too low to reproduce the abundance 
determined from observations.

Another plausible possibility to increase the HCO abundance in the PDR 
by pure gas-phase processes is to include additional reactions of 
atomic oxygen with carbon radicals that reach high 
abundances only in the PDR. Among the investigated reactions, 
the most critical one,  
\begin{equation}
\emr{O + CH_2 \rightarrow HCO + H}
\label{eq-ch2}
\end{equation}
is known to proceed with a relatively fast rate  at high temperatures
(5.01$\times$10$^{-11}$\,cm$^{3}$s$^{-1}$ at \Tk{}\,=1200-1800\,K; Tsuboi \& Hashimoto 1981). 
This is the rate recommended by NIST \cite{mal94} and UMIST2006 \cite{woo07} 
and that we adopt for our lower temperature domain ($\sim$10-200\,K).
Model predictions are shown in Fig.~\ref{fig:mods-chem} (right-side models: 
dotted curves).
Whereas the predicted HCO abundance in the shielded gas remains 
almost the same,   the HCO abundance is dramatically increased in the PDR 
(by a factor of $\sim$125) and
the O + CH$_2$ reaction becomes the HCO dominant production reaction.
Therefore, such a pure gas--phase model adding reactions 
\ref{eq-h2co} and \ref{eq-ch2}
not only reproduces the \HthCOp{} abundance in shielded core, but 
also reproduces
the observed HCO absolute abundances in the PDR.
In this picture, the enhanced HCO abundance that we observe in the Horsehead
PDR edge would be fully determined by the gas-phase chemical path :
\begin{equation}
\emr{C^+ \,\,\,\, \overrightarrow{\emr{_{H_2}}} \,\,\,\, CH_{2}^+ \,\,\,\,  \overrightarrow{_{\emr{H_2}}} 
\,\,\,\, CH_{3}^+ \,\,\,\, \overrightarrow{_{\emr{e^-}}} \,\,\,\, CH_2 
\,\,\,\, \overrightarrow{_{\emr{O}}} \,\,\,\, HCO}
\label{eq-path}
\end{equation}

 The validity of the rate of Reaction~\ref{eq-ch2} used in our PDR
model remains, of course, to be confirmed theoretically or experimentally
at the typical ISM temperatures (10 to 200\,K).

\subsection{Other routes for HCO formation: Grain photodesorption}

If Reaction~\ref{eq-ch2} is not included in the chemical network, 
the predicted HCO abundance 
is $\sim$2 orders of magnitude below the observed value towards the PDR.
As a consequence, the presence of HCO in the gas-phase should be linked to 
grain mantles formation routes, and subsequent desorption processes 
(not taken into account in our modeling).
In particular, Schilke et al. (2001) proposed that HCO could result from 
H$_2$CO photodissociation, if large quantities of formaldehyde are formed 
on grain mantles and then released in the gas phase. Even with this 
assumption, their model could not reproduce the observed HCO abundance 
in highly illuminated PDRs such as the Orion Bar. 
The weaker FUV-radiation field in the Horsehead, but large density,
prevent dust grains to acquire high temperatures over large spatial scales. 
In fact, both gas and grains  cool down below $\sim$30~K 
in $\simeq$10$''$-20$''$ (or A$_V$\,$\simeq$1-2) 
as the FUV-radiation field is attenuated. Therefore, thermal 
desorption of dust ice-mantles (presumably formed before $\sigma$-Orionis 
ignited and started to illuminate the nebula) should 
play a negligible role. Hence a non-thermal desorption mechanism
should be considered to produce the high abundance of HCO observed
in the gas phase. This mechanism could either produce  HCO directly  or 
a precursor molecule such as formaldehyde. 

Since high HCO abundances are only observed in the PDR, FUV induced
ice-mantle photo-desorption (with rates that roughly scales with the 
FUV-radiation field strength) seems the best candidate
\citep[e.g.,][]{wil93,ber95}. Laboratory experiments
 have shown that HCO radicals 
are produced in irradiated, methanol containing, ice mantles 
 \cite{bernstein95,moore01,bennett07}. 
The formyl radical could be formed through the hydrogenation 
of CO in the solid phase. It is an important intermediate radical in
the synthesis of more complex organic molecules such as methyl formate
or glycolaldehyde \cite{bennett07}. However,  the efficiency
of the production of radicals in FUV irradiated ices remains uncertain,
 and very likely depends on the ice-mantle composition. 
The formation of species like
formaldehyde and methanol in CO-ice exposed to H-atom bombardment 
 has been reported by different groups \cite{hir94,wat02,lin07}, further
confirming the importance of HCO as an intermediate product in the
synthesis of organic molecules in ices. Indeed, hydrogenation 
reactions of CO-ice,  which  form HCO, H$_2$CO, CH$_3$O and CH$_3$OH 
in grain mantles \citep[e.g.,][]{tie97,cha97}, are one important 
path which warrants further studies. 

To compare with our observations, we further need to understand 
how the radicals are released in the gas phase, either directly during the
photo-processing, or following FUV induced photo-desorption.
 Recent laboratory measurements have started to shed
light on the efficiency of photo-desorption, which depends on
the ice composition and molecule to be desorbed.  For species
such as CO, the rate of photo-desorbed molecules per FUV photon is much 
larger than previously thought \citep[e.g.,][]{obe07}. 
Similar experiments are required
to constrain the formation rate of the various species that can form 
in interstellar ices and to determine their photo-desorption rates. 

\newcommand{\kdiss}{\kappa_{\rm diss}}
\newcommand{\kpd}{\kappa_{\rm pd}}
\newcommand{\kprod}{k_{\rm prod}}
\newcommand{\phidiss}{\phi_{\rm diss}}
\newcommand{\phinth}{\phi_{\rm nth}}
\newcommand{\HCOice}{\rm HCO_{ice}}
\newcommand{\ice}{\rm H_2O_{ice}}

We can use the measured gas phase abundance of HCO to constrain
the efficiency of  photo-desorption. We assume that
the PDR is at steady state, and that the main HCO formation mechanism is
 non thermal photo-desorption from grain mantles (with a $F_{HCO}$ rate), 
while the main destruction 
mechanism is gas--phase photodissociation (with a $D_{HCO}$ rate), 
therefore :

\begin{equation}
D_{HCO} = G_0\, \kdiss(\HCO)\, \chi(\HCO)\, n(\HH) \hspace{0.5cm} [\rm cm^{-3}\,s^{-1}]
\end{equation}
\begin{equation}
F_{HCO} = G_0\, \kpd(\HCO)\, \chi(\HCOice)\,
{n(\ice) \over n(\HH)}\, n(\HH)   [\rm cm^{-3}\,s^{-1}]
\end{equation}

\noindent
where $\chi(\HCO)$ is the gas phase abundance of HCO relative to H$_2$,
$\chi(\HCOice)$ is the solid phase abundance relative to water ice, 
and $n(\ice)/n(\HH)$ is the fraction of water in the solid phase
relative to the total gas density. $\kdiss(\HCO)$ and $\kpd(\HCO)$
are the HCO photodissociation and photo-desorption rates respectively.

By equaling the formation and destruction rates, we get :

\begin{equation}
\kpd(\HCO) = \kdiss(\HCO)\, {\chi(\HCO) \over \chi(\HCOice)}\,
{n(\HH) \over n(\ice)}\hspace{0.5cm} [\rm s^{-1}]
\end{equation}
or
\begin{equation}
{\kpd(HCO) \over \rm s^{-1}}  \approx 10^{-12}\, {\kdiss(\HCO) \over 10^{-9}}\, 
                 {\chi(\HCO)/10^{-9} \over \chi(\HCOice)/10^{-2}}\,
{10^{-4}n(\HH) \over n(\ice)}
\end{equation}

\noindent
where we have used typical figures for the HCO abundance in the gas 
phase ($\sim$10$^{-9}$, see above) and solid phase 
($\sim$10$^{-2}$ see e.g. Bennet \& Kaiser 2007)
 and for the amount of oxygen present as water ice in grain mantles.


Assuming standard ISM grains with a radius of 0.1\,$\mu$m
the required photodesorption efficiency (or yield) $Y_{pd}$(\HCO):
\begin{equation}
Y_{pd}(\HCO) \simeq \frac{\kpd(\HCO)}{G_0\, exp (-2A_V)\, \pi a^2} \hspace{0.5cm} [\rm molecules\, photon^{-1}]
\end{equation}
\cite[see e.g., ][]{Dhendecourt85,ber95} 
converts to
$Y_{pd}(\HCO) \approx 10^{-4}$ molecules per photon 
(for the FUV radiation field in the Horsehead and 
A$_V$\,$\simeq$1.5, where HCO peaks). 
Therefore, the production of HCO in the gas phase
from  photo-desorption of formyl radicals could be a valid alternative
to gas phase production, if the photo-desorption
efficiency is high and HCO abundant in the ice mantles. This mechanism
also requires further laboratory and theoretical studies.

Because the formyl radical is closely related to formaldehyde and
methanol and the three species are likely to coexist in the
ice mantles, a combined analysis of the H$_2$CO, CH$_3$OH and HCO
line emissions towards the Horsehead nebula
(PDR and cores) is needed to provide more information on the
relative efficiencies of gas-phase and solid-phase routes in
 the formation  of complex organic molecules in environments 
dominated by FUV-radiation. This will be the subject of a future paper.

\section{Summary and conclusions}
\label{sec:conclusions}

We have presented interferometric and single-dish data showing the
spatial distribution  of the formyl radical rotational lines in the Horsehead
PDR and associated dense core.  The HCO emission delineates the illuminated
edge of the nebula and coincides with the PAH and hydrocarbon
emission. HCO and HCO$^+$ reach similar abundances
($\simeq$1-2\,$\times10^{-9}$) in these PDR regions where the chemistry is
dominated by the presence of FUV photons.  For the physical conditions
prevailing in the Horsehead edge, pure gas-phase chemistry is able to
reproduce the observed HCO abundances (high in the PDR, low in the shielded
core) if the O~+~CH$_2$~$\rightarrow$~HCO~+~H reaction is included in the
models.  This reaction connects the high abundance of HCO, through its
formation from carbon radicals, with the availability of C$^+$ in the PDR.

The different linewidths of HCO and H$^{13}$CO$^+$ in the line of sight
towards the ``DCO$^+$ peak'' suggest that the H$^{13}$CO$^+$ line emission
arises from the quiescent, cold and dense gas completely shielded from the
FUV radiation, whereas HCO predominantly arises from the outer surface of
the cloud (its illuminated \textit{skin}).  As a result we propose the
HCO/H$^{13}$CO$^+$ abundance ratio, and the HCO abundance itself (if
$\gtrsim$10$^{-10}$), as sensitive diagnostics of the presence of FUV
radiation fields.  In particular, regions where the HCO/H$^{13}$CO$^+$
abundance ratio (or intensity ratio if lines are optically thin) is greater
than $\simeq$1 should reflect ongoing FUV-photochemistry.

Given the rich HCO spectrum and the possibility to map its bright
millimeter line emission with interferometers, we propose HCO-H$_2$
as a very interesting molecular system for calculations of the
 \textit{ab  initio} inelastic collisional rates.

\begin{acknowledgements}
  We thank the IRAM PdBI and 30m staff for their support during the
  observations. We thank A. Faure and J. Tennyson for sending us
  the H$^{13}$CO$^+$-$e^-$ collisional rates prior to publication,
  B. Godard for useful discussions on the chemistry of carbon ions
  in the diffuse ISM, and A. Bergeat and A. Canosa for interesting
  discussions on radical-atom chemical reactions. 
  JRG is supported by a \textit{Ram\'on y Cajal} research contract
  from the Spanish MICINN and co-financed by the European Social Fund.
This research has benefitted from the financial support of the CNRS/INSU 
research programme, PCMI.
  We acknowledge the use of the JPL \citep{pic98} and Cologne 
\citep{mul01,mul05} spectroscopic
  data bases, as well as the UMIST chemical reaction data base \cite{woo07}. 
\end{acknowledgements}

\Online{} %

\begin{figure*}[t]
  \centering %
  \includegraphics[height=\hsize{},angle=270]{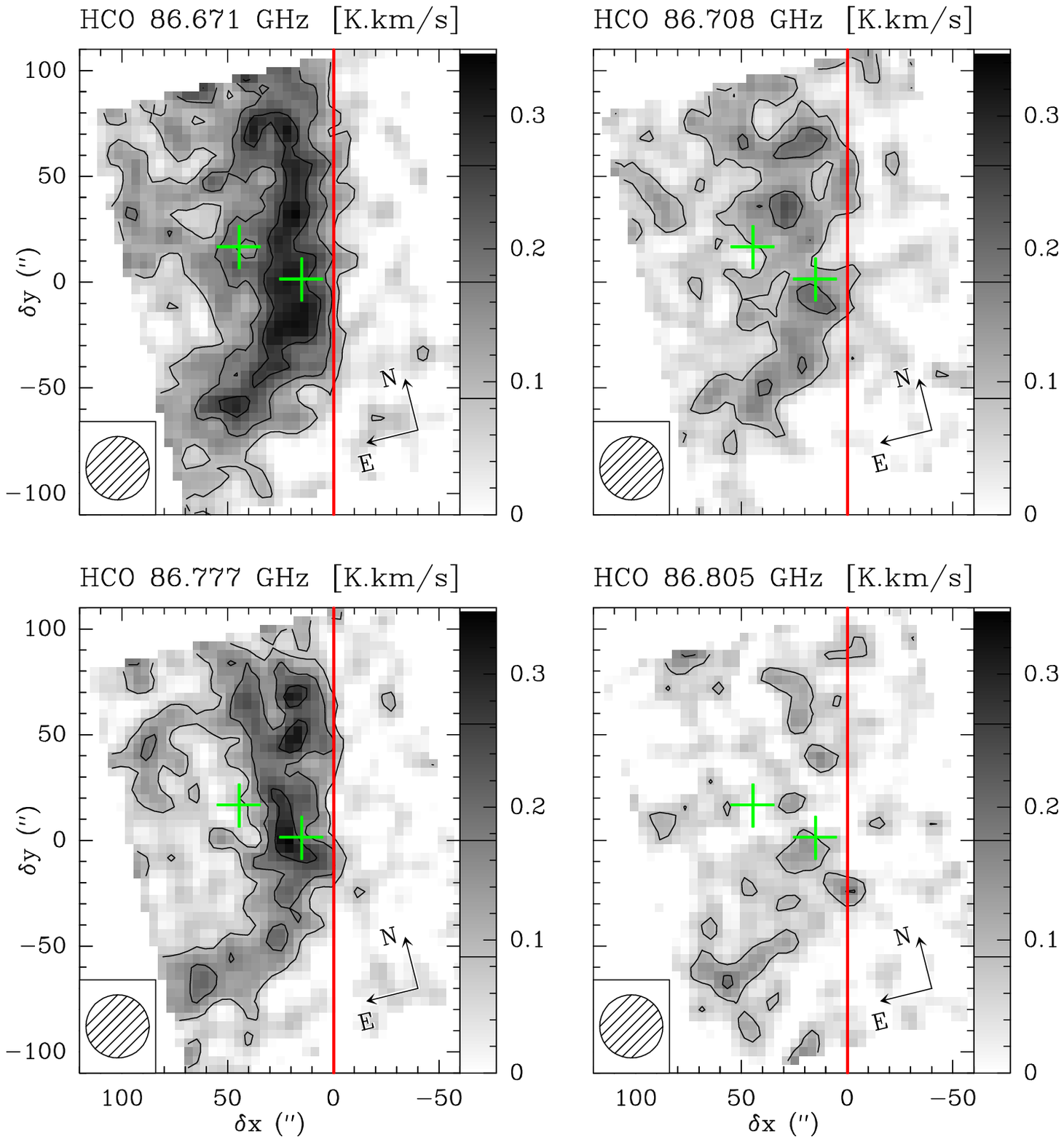} %
  \caption{Medium angular resolution maps of the integrated intensity of
    the 4 hyperfine components of the fundamental transition of HCO. 
       These lines have been
    observed simultaneously at IRAM-30m. Maps have been rotated by 14\deg{}
    counter--clockwise around the projection center, located at $(\delta
    x,\delta y)$ = $(20'',0'')$, to bring the illuminated star direction in
    the horizontal direction and the horizontal zero has been set at the
    PDR edge. The emission of all lines is integrated between 9.6 and 11.4
    \kms. Displayed integrated intensities are expressed in the main beam
    temperature scale.  Contour levels are displayed on the grey scale
    lookup tables. The red vertical line shows the PDR edge and the green
    crosses show the positions (\DCOp{} and HCO peaks) where deep
    integrations have been performed at IRAM-30m (see
    Fig.~\ref{fig:mods-30m}).}
  \label{fig:maps-30m}
\end{figure*}

\end{document}